# Synthesis of Sodium Polyhydrides at High Pressures

Viktor V. Struzhkin[1*], DuckYoung Kim[1±], Elissaios Stavrou[1], Takaki Muramatsu[1], Ho-kwang Mao[1,2], Chris J. Pickard[3], Richard J. Needs[4], Vitali B. Prakapenka[5], Alexander F. Goncharov[1]

[1]Geophysical Laboratory, Carnegie Institution of Washington,
5251 Broad Branch Road NW, Washington DC 20015, USA
[2] Center for High Pressure Science and Technology Advanced Research, Shanghai 201203, China
[3]Department of Physics and Astronomy, University College London, Gower Street, London WC1E 6BT, United Kingdom
[4]Theory of Condensed Matter Group, Cavendish Laboratory, J J Thomson Avenue, Cambridge CB3 0HE, United Kingdom
[5]Centerfor Advanced Radiation Sources, The University of Chicago, Chicago, IL 60637, USA

**Abstract:** The only known compound of sodium and hydrogen is archetypal ionic NaH. Application of high pressure is known to promote states with higher atomic coordination, but extensive searches for polyhydrides with unusual stoichiometry remain unsuccessful in spite of several theoretical predictions. Here we report the first observation of formation of polyhydrides of Na ($NaH_3$ and $NaH_7$) above 40 GPa and 2000 K. We combined synchrotron x-ray diffraction and Raman spectroscopy in a laser heated diamond anvil cell and theoretical Ab-Initio Random Structure search, which both agree in stable structures and compositions. Our results support the formation of multicenter bonding in a material with unusual stoichiometry. These results are applicable to the design of new energetic solids and high-temperature superconductors based on hydrogen-rich materials.

Dense hydrogen is of central interest in many disciplines as it is expected to possess unusual properties such as high energy density (*1*), high-temperature superconductivity and superfluidity (*2*). Metastable metallic phase of hydrogen was predicted to exist at ambient pressure, in unusual anisotropic structure, consisting of weakly interacting chains of hydrogen atoms with interatomic distances about 1.06 Å (*3*). Ashcroft (*4*) proposed that metallic state could be realized in hydrogen-rich alloys, in which hydrogen is pre-compressed by the host material, while the electronic bands of hydrogen and the host element(s) overlap. For example, metallic polyhydrides of lithium have been proposed (*5, 6*) to exist at pressures about 100 GPa, well below the expected metallization pressure of pure hydrogen (*7*). Stable polyhydrides of alkali metals have been predicted recently by theoretical analysis of $MH_n$ (M= Li, Na, K, Rb, Cs) compounds with variable hydrogen composition (*5, 6, 8-11*). The compounds with n≥2 are expected to become stable at pressures as low as 25 GPa in the case of Na and above 100 GPa in the case of Li. New polyhydride phases are also anticipated for alkaline earth metals (*10, 12, 13*). The theoretically predicted polyhydrides are not only stabilized by compression, but are also expected to metalize and exhibit superconducting properties at lower pressures than the constituent parts--the monohydride and hydrogen. Critical superconducting temperatures as high as 235 K have been predicted for polyhydrides of Ca (*12*). These compounds provide new chemical means to pre-compress hydrogen molecules and to facilitate the creation of metallic superconducting hydrogen (*14*) with record high critical superconducting temperatures. Notably, linear $H_3^-$ ions were predicted to form in polyhydrides of softer alkali metals, for example in $RbH_5$ (*11*) and $CsH_3$ (*15*). Such $H_3^-$ ions have a tendency to be stabilized in a linear configuration, contrary to triangular-shaped $H_3^+$ ions (*16, 17*). Symmetric $H_3^-$ ions with interatomic distance 1.06 Å and with the lowest potential barrier (*18*) were discussed as transition states in hydrogen exchange processes of metal complexes (*19*). $H_3^-$ and $D_3^-$ ions were experimentally observed in discharge plasmas only recently (*20*). The chains of $H_3^-$ ions that were found theoretically in $RbH_5$ (*11*) and $CsH_3$ (*15*), resemble one-dimensional hydrogen chains of ambient pressure metastable metallic hydrogen phases predicted in 1976 by Browman and Kagan (*3*). Such chains were also extensively discussed as a simplest model of strong correlations in a linear chain of hydrogen atoms (*21, 22*).

Despite a wealth of theoretically predicted high pressure polyhydride structures, none of the predictions has been confirmed until now. Here, we report synthesis of Na polyhydrides at pressure of about 30 GPa in laser heated diamond anvil cell (DAC) experiments at temperatures above 2000 K. Our *ab-initio* theoretical search yielded a number of stable $NaH_x$ (x=1.5-13) materials (Fig. 1) more favorable than those predicted previously (*8*). In agreement with these predictions, we identified the $NaH_3$ solid using *in-sit*u synchrotron x-ray diffraction measurements. Moreover, both x-ray diffraction and Raman spectroscopy revealed the presence of the $NaH_7$ phase, which has a characteristic Raman band at 3200 cm$^{-1}$, suggesting the formation of $H_3^-$ ions. Our results therefore provide the first verification of the existence of polyhydrides of alkali metals with heterogeneous (multicenter) chemical bonding and prospects for lower pressure metallization.

We have studied the formation of Li and Na polyhydrides in a DAC at pressures up to 70 GPa with laser heating to 2000 K and higher temperatures. The experiments were performed in a symmetric DAC (*23*). The samples of Li, Na, LiH, and NaH were loaded, along with small fragments of Au, in a glove box with controlled atmosphere (less than 1 ppm of oxygen). According to recent experimental (*24*) and theoretical (*25*) results, no chemical reaction is expected between Au and $H_2$, up to the highest pressure of this study. Each sample was sealed in a DAC inside a glove box, and transferred to a gas-loading apparatus, where a $H_2$ pressure of about 200 MPa was created. The DAC was opened under the $H_2$ pressure to let the gas in, resealed, and then taken out for further high-pressure experiments.

X-ray diffraction (XRD) measurements and on-line laser heating were performed at the Sector 13 (GSECARS), Advanced Photon Source at the Argonne National Laboratory(*26*) . Several experiments were performed with Li and Na samples up to 70 GPa at room temperature. In these runs only the formation of LiH and NaH was detected, with no indication of polyhydride phases. The results of these experiments were similar to previously reported attempts (*27*), however, we were able to identify Li and Na metals up to 35 GPa, and 50 GPa, respectively, without complete transformation to the monohydride form. To overcome possible kinetic barriers to the formation of polyhydrides, we performed laser-heating experiments on these samples. For Li in hydrogen we were able to perform a few experiments above 50 GPa with laser heating up to 1900-2000 K, in which only the monohydride of Li (LiH) was formed. We were not able to detect any polyhydrides of Li under these conditions. Similar measurements for Na in hydrogen at 32 GPa yielded a significant enhancement of the XRD signal from NaH. Further heating of Na and NaH in $H_2$ saturated environment to ~ 2100 K produced a laser flash that resulted in sample morphology changes, indicating the onset of chemical reactions. The Raman spectra collected from temperature quenched sample within reacted area showed the formation of a new material with two additional vibron peaks around 4000 $cm^{-1}$, one of them softer than the pure $H_2$ vibron, and another one harder (Fig. 2). However, we were unable to detect a reliable XRD signal from the very tiny sample reaction area. We repeated the laser heating experiment with a NaH sample loaded in the DAC with $H_2$ and Au fragments for measuring pressure and for better coupling to the laser heating. This experiment produced large amounts of a new phase after laser heating at 30 GPa. As the temperature was increased above 2100 K the heating conditions ran away and resulted in a very bright flash (avalanche) saturating the detector. From the brightness of the heating spot, we estimated the temperature to be in the range of 4000-6000K. We did not attempt to repeat heating due to the risk of breaking the diamonds but saved the sample for further characterization. After heating we could clearly see the change in the sample shape, indicating the sample transport within the laser-heated reaction area of around 20 μm in diameter.

The newly synthesized phases were characterized by XRD and Raman measurements in the pressure range from 18 to 50 GPa. Decompression of the DAC below 18 GPa resulted in decomposition of the newly formed phase, which was confirmed by the disappearance of their characteristic Raman signatures. These experiments are very challenging since the presence of hydrogen under high pressure-temperature conditions often leads to diamond breakage. Most of the experiments resulted in failure of the diamond during laser heating; however, we succeeded in producing Na polyhydrides in two runs out of ten, and characterized them using Raman spectroscopy and XRD. The experimental results are described below. Before describing these results, we summarize our theoretical findings which differ in a number of aspects from the previous theoretical study of Baettig *et al*. (*8*). These differences are crucial for understanding our experimental findings.

We performed searches for low enthalpy structures using a variety of compositions of Na-H at 50 GPa and the *Ab-Initio* Random Structure Searching (AIRSS) method (*6*), which has previously been applied to hydrides under pressure(*28*),(*6*). All calculations were performed using the Perdew-Burke-Ernzerhof density functional (*29*). We also performed calculations for the structures reported by Baettig *et al*. (*8*) and successfully reproduced their data for $NaH_7$, $NaH_9$, and $NaH_{11}$. We used AIRSS to study other compositions, that led to the discovery of the $NaH_3$ phase. This prompted us to extend our searches to lower hydrogen compositions such as $NaH_2$, $Na_3H_5$, and $Na_2H_3$. For most compositions, we studied simulation cells containing 1,2, and 4 formula units (f. u.), and for $NaH_2$ and $NaH_3$, we conducted AIRSS on up to 6 f. u. The most stable materials found consisted of $H_2$ and NaH structural units. This finding leads us to generalize the form of the stable composition to $(NaH)_m(H_2)_n$ (Fig. 1). We studied ($m,n$) pairs ranging from (4,1) to (1,6). We also tested other compositions such as $Na_2H_5$, $Na_2H_7$, and $Na_2H_9$, but we found them unstable with respect to decomposition into nearby stable compositions, as shown in Fig.1. Previous theoretical work suggested that $NaH_n$ (n > 6) can be stabilized above 50 GPa (*8*). As shown in the convex hull diagram of Fig. 1 at 50 GPa, generally, any combination of (NaH) and $H_2$ can be stabilized. The $Na_2H_3$, $Na_3H_5$, $NaH_3$, $NaH_9$, and $NaH_{13}$ phases shown in blue lie on the convex hull at 50 GPa. In addition, although they are not thermodynamically stable, $NaH_2$, $NaH_5$, $NaH_7$, and $NaH_{11}$ (shown in green) are dynamically stable as demonstrated by the phonon dispersion data reported in the Supplementary Materials and in the paper of Baettig *et al*. (*8*).
The enthalpy differences between the thermodynamically and dynamically stable phases (blue line) and the dynamically stable phases (green) is only around 10 meV/atom. We therefore calculated the nuclear zero point energy (ZPE) within the harmonic approximation to estimate the effects of vibrations on the total enthalpy. We found a monotonic increase in the ZPE with the fraction of H atoms in the various hydrides, ranging from 150 meV / atom in NaH to ~ 240 meV / atom in $H_2$ (Supplementary Materials, Fig.S31).

In the Supplementary Materials, we report the convex hull including ZPE effects, which shows that NaH$_7$ comes within 1~2 meV/atom of being thermodynamically stable (Fig.S32). The changes to the convex hull of Fig. 1 from including the ZPE are small, although the NaH$_7$ structure moves to within 1~2 meV/atom of being thermodynamically stable.

The Raman spectra of the NaH$_n$ materials synthesized by laser heating (Fig. 2) show a number of features, which are distinct from those of the pure hydrogen within the same sample chamber under the same pressure (50 GPa). New modes at 4100 cm$^{-1}$ and 4200 cm$^{-1}$, bracketing the H$_2$ vibron at 4160 cm$^{-1}$, point to the formation of a new phase containing H$_2$ molecules embedded within the sodium polyhydrite crystal structure. Moreover, as shown in Fig. 2, another set of Raman modes appears around 3200 cm$^{-1}$, suggesting a strongly modified H$_2$ species, possibly similar to the predicted H$_3^-$ species or molecules in polyhydrides of Cs (*11*) or K (*9*). Similar or even lower Raman frequencies are characteristic of dihydrogen moieties observed in transition metal complexes (*30*),(*31*) and other chemical environments (*32*). The low frequency regions of the Raman spectra (Supplementary Materials, Fig. S3) also suggest a structure very different from pure hydrogen (e.g.(*33*)) and the initial body centered cubic (bcc) NaH monohydride, which is not expected to have any allowed first order Raman active modes. Indeed, our Raman measurements for unreacted sample regions in the DAC did not produce any Raman signatures of NaH, but indicated the presence of pure solid H$_2$ from its characteristic vibron and roton bands. The low-frequency Raman spectrum of the newly synthesized material consists of strongly pressure-dependent bands at 200-800 cm$^{-1}$, which we identify as lattice modes in contrast to weakly pressure-dependent rotational modes of pure H$_2$. (Fig. S3 of the Supplementary Materials)

Fig. 3 shows an XRD pattern of a new material at 40 GPa. XRD data were also obtained away from the reacted area at each pressure (see inset to Fig S1 of the Supplementary Materials). Three different "families" of reflections from different phases were observed to coexist in the XRD patterns of the reacted area: i) the unreacted bcc NaH [ambient pressure face-centered cubic (fcc) NaH transforms to bcc at 29 GPa (*34*)], ii) the fcc Au used as a pressure marker and as laser absorber, and iii) the synthesized NaH$_n$. In order to fully identify the reflections from the synthesized NaH$_n$ we performed a detailed comparison of the XRD patterns on and away from the reacted area. A typical example is shown in Fig. S1 of the Supplementary Materials. The positions of all reflections attributed to NaH and Au, are in full agreement with the known EOS of bcc NaH (*34*) and fcc Au, implying the absence of chemical reaction between Au and H. The reflections of bcc NaH and Au have then been subtracted when performing the final structural refinement of the NaH$_n$ phases. This has been performed via a Rietveld refinement only for bcc NaH and fcc Au with a subsequent subtraction of the refined peaks from the raw patterns. A typical example of this procedure is shown in Fig. S2 of the Supplementary Materials. After all reflections not belonging to the synthesized NaH$_n$ being successfully identified we compared the calculated XRD patterns of the predicted stable structures with the observed ones. Full indexing-refinement of the observed reflections, without the use of the predicted phases as candidates, is very difficult for a variety of reasons. First, the XRD intensity depends almost exclusively on the positions of the Na atoms. Second, the large number of observed peaks suggests a low symmetry unit cell. Finally, the texture of the 2-D images of the XRD data suggests a mixture of phases. Based on this analysis, we find that NaH$_3$ is the predominant phase of the synthesized material (Fig. 3). Indeed, all the main reflections can be indexed with the orthorhombic *Cmcm* NaH$_3$ cell. Moreover, the experimentally determined lattice parameters and cell volume (at 40 GPa: a=3.332 Å, b= 6.354 Å and c= 4.142 Å with V$_{pfu}$=21.93 Å$^3$) of NaH$_3$ are in full agreement with the theoretical predictions (Fig. 4). However, there are a few reflections that cannot be indexed with the NaH$_3$ cell. For hydrogen contents lower than in NaH$_5$, the phonon density of states has two well-separated bands, below 1500 cm$^{-1}$ for Na-H interactions and around 4000 cm$^{-1}$ for H$_2$ vibrations. At higher hydrogen concentrations, we found the formation of other intermediate frequency bands near 3200 cm$^{-1}$. Having in mind that NaH$_n$ phases (n <7) cannot support the existence of Raman modes at 3200 cm$^{-1}$ (Supplementary materials) we have to include phases with n>6 (7, 9) in our analysis. From the various phases only the monoclinic *Cc* NaH$_7$ phase shows reasonable agreement with the observed patterns. Indeed, some of the main observed reflections can only be indexed with the NaH$_7$ phase with lattice parameters a=6.732 Å, b=3.643 Å, c= 5.577 Å and *β*=69.36° at 40 GPa. With the use of both phases, $NaH_3$ and $NaH_7$, we have successfully indexed all observed reflections of the synthesized mixed-$NaH_n$ material (Fig. 3). There is a very good agreement between observed and theoretically predicted relative intensities of Bragg reflections. However, a refinement of the positional parameters was not possible due to the "spotty" XRD rings.

The Raman and XRD data point to the formation of Na polyhydrides in the predicted stability range (above 20 GPa). While we were unable to isolate a single well defined polyhydride phase, the data analysis strongly supports the existence of several phases ($NaH_3$ and $NaH_7$, and possibly higher polyhydrides) in the reacted sample. Most of the theoretically predicted stable Na polyhydride phases have low symmetry structures, which are extremely difficult to characterize by XRD from the small samples available in the laser-heated region. While prolonged laser heating at well-defined *P-T* conditions may be beneficial for growing a single-phase sample, such experiments are still inaccessible due to the high reactivity of hot hydrogen with diamond anvils. Notably, Raman spectroscopy provided a more sensitive tool than XRD for characterizing formation of small amounts of low-Z polyhydride materials. Based on the results of theoretical calculations (Supplementary Materials), we found that the Raman bands observed experimentally near 3200 cm$^{-1}$ can be assigned to an extended hydrogen molecular $H_2$ unit with an intramolecular length *d* of ~ 0.81 Å. This $H_2$ molecule is linked to a hydrogen atom in the NaH unit with a distance of *z*= 1.25 Å, and they form a $H_3^-$ linear anion in $NaH_x$ materials with x=7 (Fig. 2 and Supplementary Materials). It was suggested that pressure can induce a linear geometry for $H_3^-$, which has four electrons, but a triangular geometry for $H_3^+$ which has two electrons([17]) and a careful theoretical study[5] of heavy alkali-metal hydrides under pressure predicted the formation of linear $H_3^-$ in $KH_5$. To gain further insights into $H_3^-$ anion formation in $NaH_7$, we analyzed the charge density of $NaH_3$ and $NaH_7$ using Bader method. The calculations confirmed the highly ionic nature of the NaH unit in each polyhydrides: the net charges on Na and H in the NaH unit are + 0.79/0.82 and -0.65/0.47 in $NaH_3$/$NaH_7$, respectively, indicating that a portion of the electrons is donated to the $H_2$ molecules in $NaH_7$. In fact, the $H_3^-$ anion in $NaH_7$ has an excess of -0.63 electrons which leads to elongation of the $H_2$ bond.

*Ab-initio* phonon calculations give information on the dynamical stability of the phases. The stability region of NaH$_7$ was predicted (*8*) to be 25-100 GPa which is consistent with our experiments. All lattice and vibron modes of the polyhydrides increase monotonically in frequency with pressure up to 50 GPa. Our theoretical calculations show dynamical stability and structural stability of predicted phases, including NaH$_3$ and NaH$_7$.

In summary, we have synthesized polyhydrides of Na in a laser-heated DAC at pressures above 30 GPa and temperatures above 2000 K. We also have performed detailed theoretical studies and found new stable phases of Na polyhydrides. One of these phases, NaH$_3$, matches well with the XRD patterns collected from the reacted region. However, the x-ray patterns also suggest the existence of higher polyhydrydes (NaH$_n$, n$\geq$7), which is supported by analysis of the Raman spectra in the 3200 cm$^{-1}$ region. Notably, higher polyhydrides of sodium appear to stabilize H$_3^-$ unit predicted for other, softer alkali metals (*15*). Polyhydrides of alkali metals open a new class of materials with pressure-stabilized multicenter (3 center – 4 electron) bonds for future investigation. Polyhydrides may provide chemical means to pre-compress hydrogen molecules and facilitate the creation of metallic superconducting hydrogen at reduced pressures. The possibility of metastable phases should be carefully explored in the future studies, since the new polyhydrides may be implemented as hydrogen storage materials with hitherto unexplored physical and chemical properties.

**Acknowledgements:**

High pressure experiments were supported by DOE/BES under contract No. DE-FG02-02ER45955. Theoretical analysis (D. Y. K) was supported by Energy Frontier Research in Extreme Environments Center (EFree**)**, an Energy Frontier Research Center funded by the U.S. Department of Energy, Office of Science under Award Number DE-SC0001057**.** C. J. Pickard and R. J. Needs were supported by the Engineering and Physical Sciences Research Council (EPSRC) of the UK. E. Stavrou and A. F. Goncharov acknowledge support of DARPA.

Portions of this work were performed at GeoSoilEnviroCARS (Sector 13), Advanced Photon Source (APS), Argonne National Laboratory. GeoSoilEnviroCARS is supported by the National Science Foundation - Earth Sciences (EAR-1128799) and Department of Energy - Geosciences (DE-FG02-94ER14466). Use of the Advanced Photon Source was supported by the U. S. Department of Energy, Office of Science, Office of Basic Energy Sciences, under Contract No. DE-AC02-06CH11357.

We thank K. Zhuravlev and S. Tkachev for help with x-ray diffraction and Raman measurements at GSECARS.

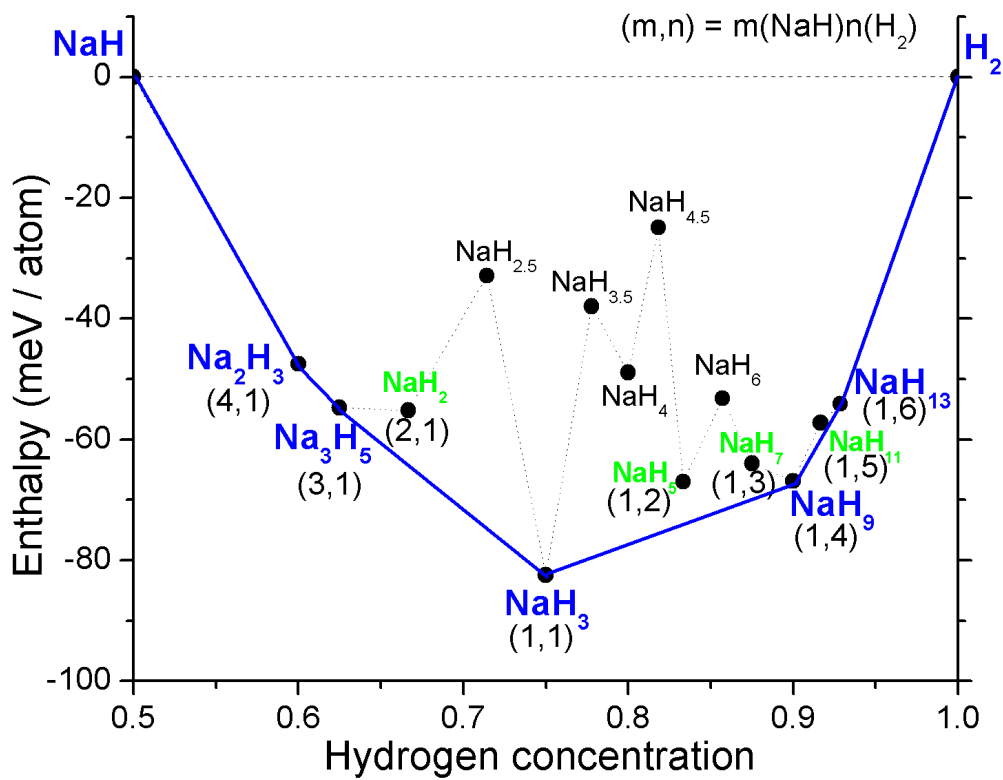

**Fig. 1.** Convex hull curve of Na-H compounds at 50 GPa with respect to the decomposition (horizontal dashed line) into NaH and $H_2$. Chemical formula in blue (green) shows predicted stable (meta-stable) compounds. The (m,n) correspond to compositions in units of NaH and $H_2$, respectively.

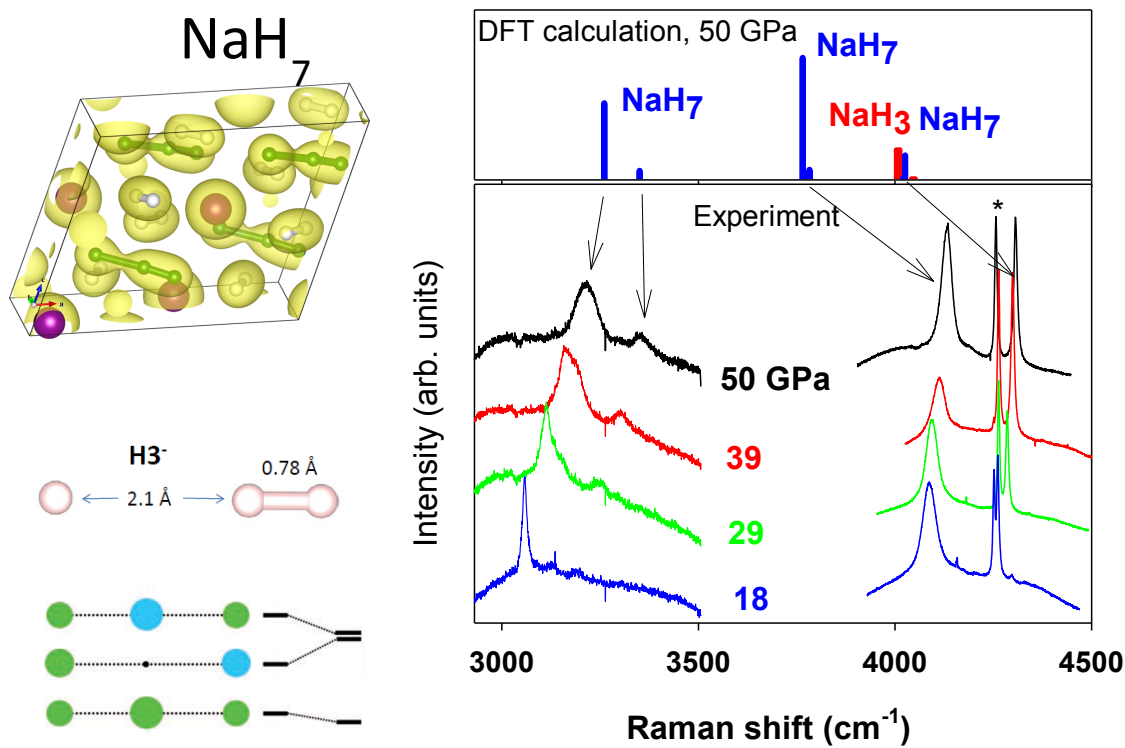

**Fig. 2.** Raman spectra of NaH$_n$ sample, showing higher-frequency vibrons from H$_2$ molecular-type structural units.

The left panel shows the structure of NaH$_7$, which contains H$_3$ complexes. The schematic diagram for H$_3$- molecular orbitals from Ref. (*17*) is also shown.

Right panel: Raman spectra of NaH$_7$ sample in the frequency region typical for vibrons from H$_3$- units (indicated in the structure of NaH$_7$ as green-yellow dumbells) are shown in 3000-3500 wavenumber region. The top panel shows calculated Raman Intensity for NaH$_3$ and NaH$_7$. Raman from pure H2 vibron is indicated by asterisk.

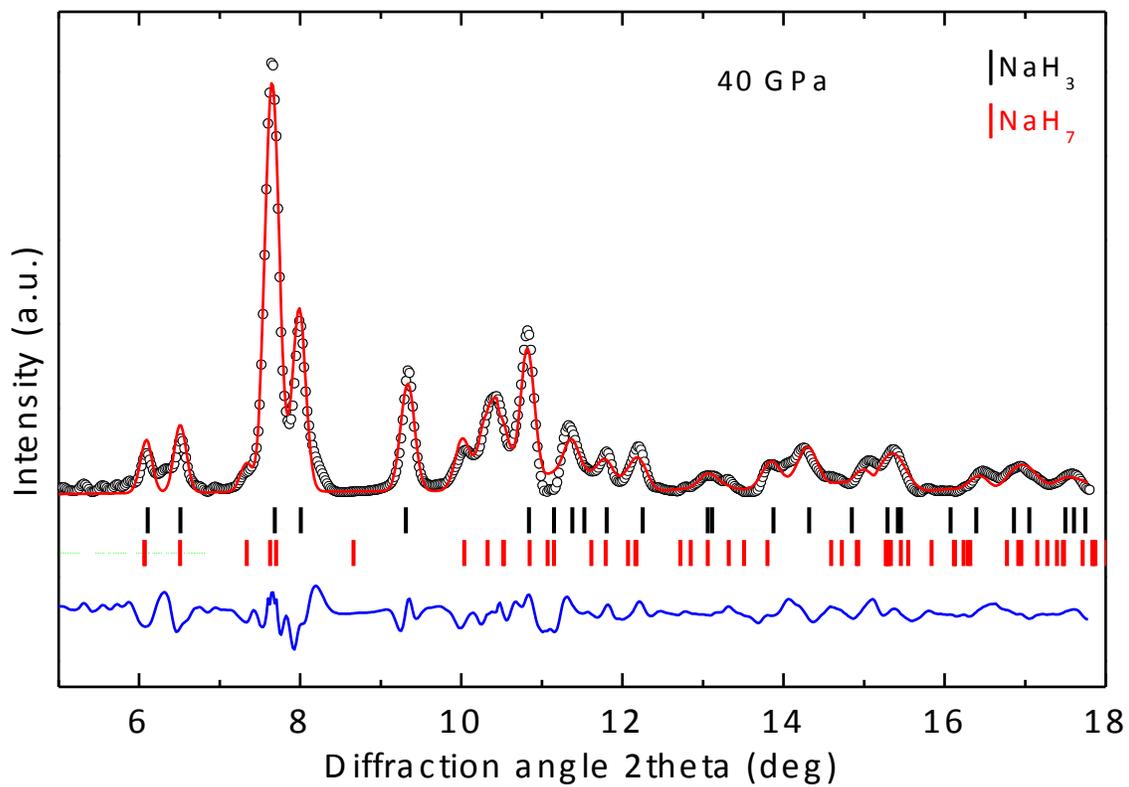

**Fig. 3**. Le Bail refinement for NaHn at 40 GPa. NaH$_3$ and NaH$_7$ peaks are marked with black and red vertical lines respectively.

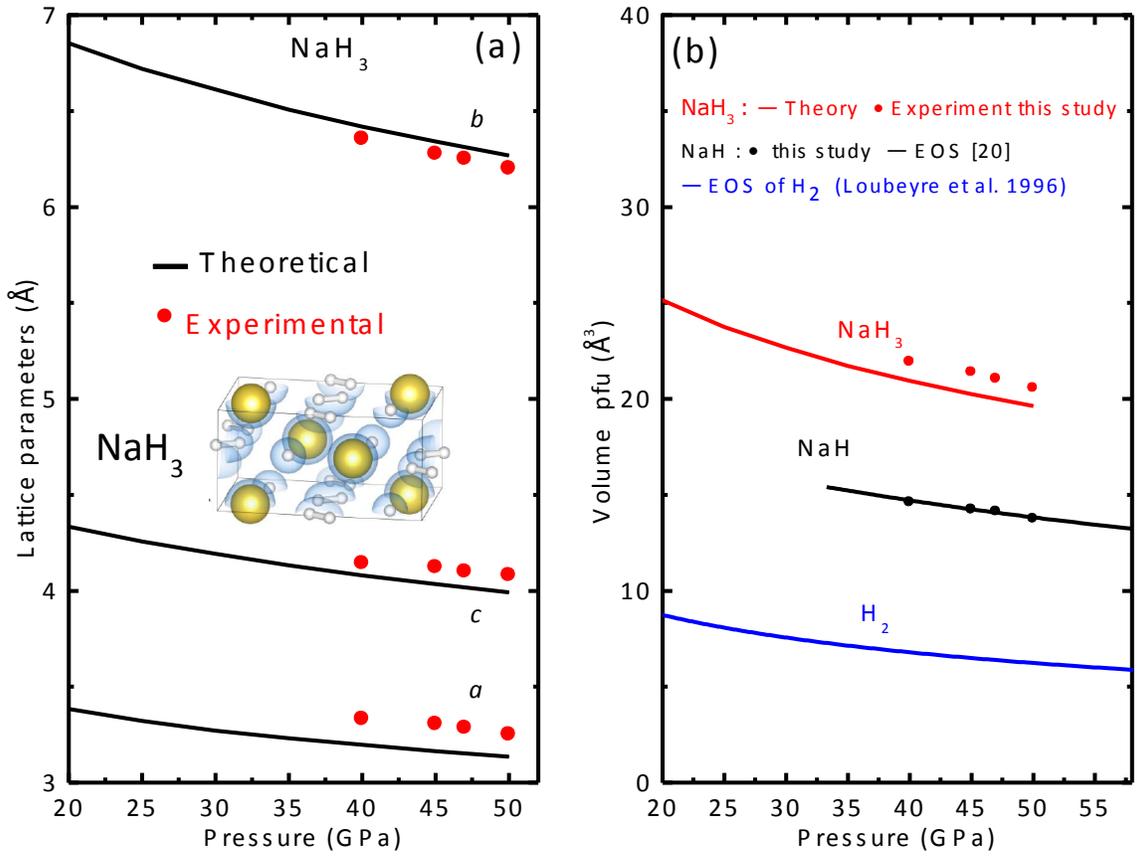

**Fig. 4**. (a) Lattice parameters of NaH$_3$ as function of pressure.
(b) EOS of NaH$_3$ in comparison with EOS of NaH and H$_2$.
Experimental data: red and black circles, theoretical predictions: red and black continuous lines. EOS of H$_2$ is also shown with a blue line.
The structure of NaH$_3$ is shown in the panel (a).